# Ab initio spectroscopic characterization of the radical $CH_3$-O-$CH_2$ at low temperatures

**O. Yazidi**

Laboratoire de Spectroscopie Atomique Moléculaire et Applications, Faculté des Sciences de Tunis, Université de Tunis El Manar, 2092 Tunisia.

**M. L. Senent[a] and V. Gámez**

Departamento de Química y Física Teóricas, Instituto de Estructura de la Materia, IEM-CSIC, Serrano 121, Madrid 28006, Spain, Unidad Asociada GIFMAN, CSIC-UHU; 21071 Huelva, Spain.

**M. Carvajal**

Dpto. Ciencias Integradas, Centro de Estudios Avanzados en Física, Matemática y Computación, Facultad de Ciencias Experimentales, Universidad de Huelva; Unidad Asociada GIFMAN, CSIC-UHU; 21071 Huelva, Spain.  Instituto Universitario Carlos I de Física Teórica y Computacional, Universidad de
Granada, Granada, Spain

**M. Mogren Al-Mogren**

King Saud University, Chemistry Department,
Faculty of Science, P.O. Box 2455,
Riyadh 11451, Kingdom of Saudi Arabia.

---

**ABSTRACT**

Spectroscopic and structural properties of methoxymethyl radical ($CH_3OCH_2$, RDME) are determined using explicitly correlated ab initio methods. This radical of astrophysical and atmospheric relevance has not been fully characterized at low temperatures, which has delayed the astrophysical searches. We provide rovibrational parameters, excitations to the low energy electronic states, torsional and inversion barriers and low vibrational energy levels. In the electronic ground state ($X^2A$), which appears "clean" from non-adiabatic effects, the minimum energy structure is an asymmetric geometry which rotational constants and dipole moment have been determined to be $A_0$=46718.6745 MHz, $B_0$=10748.4182 MHz, and $C_0$=9272.5105 MHz, and 1.432 D ($\mu_A$=0.6952D, $\mu_B$=1.215D, $\mu_C$=0.3016D), respectively. A variational procedure has been applied to determine torsion-inversion energy levels. Each level splits into 3 subcomponents ($A_1/A_2$ and E) corresponding to the three methyl torsion minima. Although the potential energy surface presents 12 minima, at low temperatures, the infrared band shapes correspond to a surface with only three minima because the top of the inversion $V^\alpha$ barrier at $\alpha$=0º (109 cm$^{-1}$) stands below the zero point vibrational energy and the $CH_2$ torsional barrier is relatively high (~2000 cm$^{-1}$). The methyl torsion barrier was computed to be ~500 cm$^{-1}$ and produces a splitting of 0.01 cm$^{-1}$ of the ground vibrational state.

**INTRODUCTION**

Radicals play important roles in the atmospheric and in the interstellar chemistry. In gas-phase sources and grains of the interstellar medium, they can induce exothermic and no activation energy processes [1-3]. In the atmosphere, most of the chemical reactions involve reactive free radicals that are generated by photochemical processes from stable precursors. Methoxymethyl radical ($CH_3OCH_2$, RDME) is produced as a result of removing one methyl hydrogen from dimethyl ether ($CH_3OCH_3$, DME). Atmospheric degradation of DME can be initiated via hydrogen subtraction by OH radicals, molecular oxygen or atomic hydrogen to produce RDME [4-6]. Experimental and theoretical studies aiming to clarify the mechanisms of DME oxidation in the atmosphere are recurrent [7-18]. Since DME is an abundant interstellar molecule, the H subtraction processes have been studied at very low temperatures obtaining RDME as a product [1].

The application of dimethylether as an alternative fuel has motivated frequent previous kinetic studies where RDME appears as a common product [8]. However, spectroscopic studies are quite limited. The UV absorption spectrum of the $CH_3OCH_2$ radical was recorded by Langer et al. [19] in the gas phase. The pulse radiolysis of a mixture containing DME led to a rapid increase in the UV absorption at 230 nm associated to RDME. Broad bands were observed between 3.5-5.6 eV (1 eV=1.602176 x$10^{-19}$ J) [19]. Vertical excitations energies computed using multireference configuration interaction theory (MRCI) and a double zeta basis set [20] allowed to identify the excitation to the first excited electronic state with the band observed at 4.13 eV [19]. Four states with doublet spin-multiplicity character were predicted at 4.03, 5.16, 6.83, and 7.44 eV [20].

To date, there are not available measurements of the rotational spectra of the methoxymethyl radical. In fact, to our knowledge, there is only a unique published paper addressing the infrared spectrum [21]. Recently, Gong and Andrews have recorded the IR spectrum in Ar matrix [21] characterizing four infrared absorptions at 1468.1, 1253.9, 1226.6, and 944.4 cm$^{-1}$, which were assigned by deuterium substitution as well as, by frequency and intensity calculations using density functional theory.

In astrophysical models, RDME is considered a possible intermediate of processes connecting the two abundant species DME and methyl formate [3] and it is considered a detectable species. The discovery of new molecules requires a previous laboratory characterization that, in our case, involves intricate experiments due to the high reactivity of radicals. The lack of available spectroscopic parameters and the enormous astrophysical interest are the motivation of the present paper which aims to obtain as much as possible information that can be derived from state-of-the-art ab initio calculations. We search to determine accurate theoretical parameters for further spectral assignments and further astrophysical searches. Highly correlated ab initio methods are employed to obtain reliable rovibrational parameters (rotational constants, centrifugal distortion constants, vibrational band centers,..), the dipole moment, excitation to the low excited electronic states, and an exhaustive description of the far infrared spectral region.

RDME can be defined as non-rigid species because the ground electronic state potential energy surface presents twelve minima connected by large amplitude motions. Three internal large amplitude modes are responsible for the non-rigidity: the methyl group and the $CH_2$ group torsions and the $CH_2$ inversion. Because the methyl torsion barrier and the inversion barrier are much lower than the $CH_2$ torsional barrier, in principle, the low energy levels can be calculated using a variational two-dimensional model. This allows us to obtain splittings of the levels due to tunneling effect.

# COMPUTATIONAL DETAILS

The equilibrium structure and the two-dimensional potential energy surface of RDME were calculated with explicitly correlated coupled cluster theory, RCCSD(T)-F12b [22-23] implemented in MOLPRO [24] using the corresponding default options. Furthermore, a full-dimensional anharmonic force field was computed with second order Möller-Plesset theory (MP2) implemented in GAUSSIAN [25]. This force field was applied to determine the vibrational corrections of the surface and the anharmonic contributions [26] which are less dependent on the level of theory than the first order spectroscopic properties. The aug-cc-pVTZ basis set (denoted by AVTZ) [27] was used in the MP2 calculations. For the RCCSD(T)-F12 calculations, the AVTZ atomic orbitals were employed in connection with the corresponding basis sets for the density fitting and the resolution of the identity.

Vertical excitation energies to the excited electronic states were determined with the complete active space self-consistent field (CASSCF) method [28-29] followed by the internally contracted multi reference configuration interaction approach (MRCI) [30-31]. Both methods are implemented in MOLPRO [24].

For the two modes responsible for the non-rigidity being the methyl torsion and the $CH_2$ wagging (or inversion) modes, the energies were calculated using the original program ENEDIM [32] and a variational model of reduced dimensionality. More details concerning the theory implemented in ENEDIM theory, as well as examples of previous applications, can be found in Ref. [33-38].

## RESULTS AND DISCUSSION

**Electronic ground state: molecular structure and rotational parameters**

The most stable structure of RDME shown in Figure 1 is an asymmetric geometry that can be classified in the $C_1$ point group. Three large amplitude motions, the $CH_3$ and the $CH_2$ internal rotation, and the $CH_2$ wagging, intertransform the 12 minima of the electronic ground state potential energy surface. Since the splitting of the $CH_2$ torsion is not going to have any effect in the calculations of the present work given the height of the corresponding barrier, the far infrared spectrum can be explored using a two-dimensional model and the levels can be classified using the Molecular Symmetry Group $G_6$ [39-40]. Thus, in this paper, the $CH_3$ torsion and the $CH_2$ wagging are treated as large amplitude vibrations responsible for the minimum intertransformation, whereas the $CH_2$ torsional motion is described by small displacements around the equilibrium position ($CH_2$ twist). Two symbols $\theta$ and $\alpha$ identify the corresponding large amplitude coordinates. $\alpha$ represents the dihedral angle between the C3O1C2 and H7C2H8 planes whereas the methyl torsional coordinate is defined using three dihedral angles:

$$\theta = (H4C3O1C2 + H5C3O1C2 + H6C3O1C2)/3 \qquad (1)$$

At the non-planar equilibrium structure $\theta^{MIN} = 177.1°$ and $\alpha^{MIN} = 25°$.

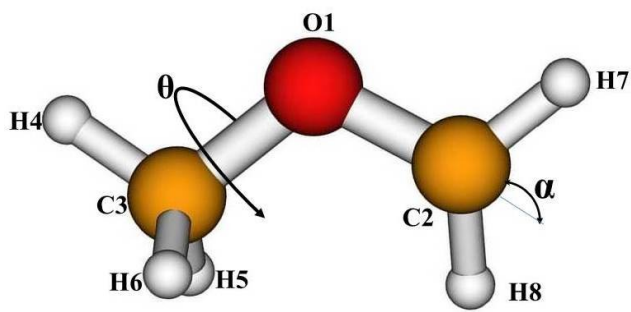

Figure 1: The minimum energy geometry of RDME. Independent coordinates and atom labelling

Table I collects the structural parameters and the equilibrium rotational constants computed at the RCCSD(T)-F12 level of theory. The dipole moment components were obtained using MP2 calculations.

**Table I:** RCCSD(T)-F12b/AVTZ structural and rotational constants. MP2/AVTZ centrifugal distortions constants and dipole moment of the minimum energy geometry of $CH_3OCH_2$.

*Structural parameters (Å, degrees)[a]*

| | | | |
|---|---|---|---|
| O1C2 | 1.3535 | H5C3O1 | 110.4 |
| O1C3 | 1.4192 | H6C3C1 | 110.4 |
| H4C3 | 1.0866 | H7C2O1 | 114.0 |
| H5C3 | 1.0936 | H8C2O1 | 118.2 |
| H6C3 | 1.0924 | H4C3O1C2 | -177.0 |
| H7C2 | 1.0782 | H5C3O1C2 | -57.7 |
| H8C2 | 1.0838 | H6C3O1C2 | 63.5 |
| C2O1C3 | 115.4 | H7C2O1C3 | 176.9 |
| H4C3O1 | 107.0 | $\alpha^{MIN}$ | 25.0 |

*Rotational constants (MHz)*

| | | | |
|---|---|---|---|
| $A_e$ | 46625.72 | $A_0$ | 46718.67 |
| $B_e$ | 10859.71 | $B_0$ | 10748.42 |
| $C_e$ | 9367.67 | $C_0$ | 9272.51 |

*Centrifugal distortions constants (S reduction, $I^r$ representation )*

| | | | |
|---|---|---|---|
| $\Delta_J$ (MHz) | 0.0103 | $H_J$ (Hz) | 0.0008 |
| $\Delta_{JK}$ (MHz) | -0.0357 | $H_K$ (Hz) | 27.7329 |
| $\Delta_K$ (MHz) | 0.5832 | $H_{JK}$ (Hz) | -1.0954 |
| $d_1$ (MHz) | -0.0021 | $H_{KJ}$ (Hz) | -5.8508 |
| $d_2$ (MHz) | 0.0002 | $h_1$ (Hz) | 0.0041 |
| | | $h_2$ (Hz) | 0.0029 |
| | | $h_3$ (Hz) | 0.0007 |

*Dipole moment (D)[a]*

| | | | | | | | |
|---|---|---|---|---|---|---|---|
| $\mu$ | 1.432 | $\mu_A$ | 0.695 | $\mu_B$ | 1.215 | $\mu_C$ | 0.302 |

a) 1Å=$10^{-10}$ m; 1D =3,33564×$10^{-30}$ Coulombs x m.

Table I shows the ground vibrational state rotational parameters corresponding to the Watson S-reduction Hamiltonian ($I^r$ representation) [41]. The centrifugal distortions constants were computed from an anharmonic MP2 force field. The ground vibrational state rotational constants $A_0$, $B_0$ and $C_0$ were estimated from the corresponding

RCCSD(T)-F12 equilibrium parameters ($A_e$, $B_e$ and $C_e$), using the following equation [36,42-44],

$$B_0 = B_e \text{(RCCSD(T)-F12)} + \Delta B_e^{core} \text{(RCCSD(T))} + \Delta B^{vib} \text{(MP2)} \quad (2)$$

where $\Delta B^{vib}$ represents the vibrational contribution to the rotational constants derived from the Vibrational Second Order Perturbation Theory (VPT2), $\alpha_r^i$ vibration-rotation interaction parameters determined with the MP2 cubic force field [45], and $\Delta B_e^{core}$ contains the effect of the core-electron correlation.

Although, we have not found any available experimental data to assess the present computed rotational parameters, we can expect divergences lower than 10 MHz with the real values. This threshold is sustained by former results obtained for diverse molecular. Examples are those described in Ref. [36,42-44]. Some of these previous works, such us, the study of 4-hydroxy-2-butynenitrile [42] and dimethylsulfoxide [44] were performed in collaboration of rotational spectroscopy laboratories. On the basis of the previous knowledge, we can assert that the predicted rotational constants $A_0$=46718.6745 MHz, $B_0$=10748.4182 MHz, and $C_0$=9272.5105 MHz are accurate enough for being employed in further spectrum assignments.

**Excited electronic states**

Vertical excitation energies to the low excited electronic states were computed at the MRCI/AVTZ level of theory to explore the density of states in the ground state region and to evaluate the risk of vibronic effects due to the radical character of RDME. The resulting energy levels are shown in Figure 2 where they are classified using the irreducible representations of group $C_1$ of the minimum energy geometry ($\alpha=\alpha^{MIN}$) and the group $C_s$ of the planar structure corresponding to the top of the $CH_2$ wagging barrier ($\alpha=0°$). Both structures are very close in energy (~100 cm$^{-1}$). The ground electronic state appears "clean" from non-adiabatic effects.

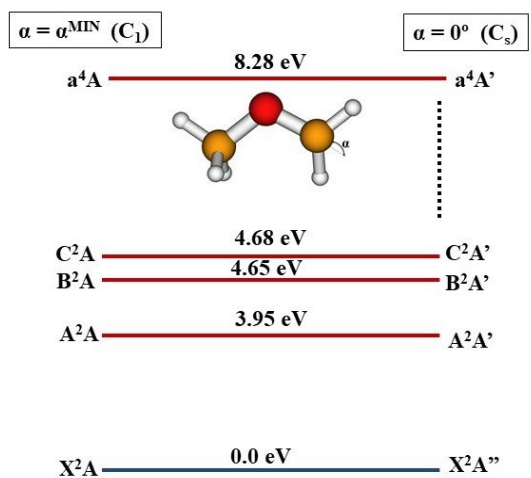

Figure 2: First four doublet electronic states of RDME.

Configuration Interaction (CI) calculations were performed over an active space of 10 orbitals, larger than the 7 orbital space employed in Ref. [20]. Nine orbitals were optimized and considered double occupied in all the configurations. The first four electronic states show doublet spin multiplicity, whereas the first quartet state appears over 8 eV.

The X$^2$A ground electronic state, as well as the low energy excited states, presents a doublet spin-multiplicity character. The first excited state (A$^2$A) appears at 3.95 eV in a region where a UV broad band has been observed (4.03 eV [19]). This excitation energy is in a good agreement with previous calculations (4.13 eV [20]). High density of states characterizes the 3.9-4.7 eV region, as it has been experimentally observed [19].

**Infrared spectrum**

The fundamental transitions of Table II were estimated using the following formula:

$$E=\sum_i \omega_i^{CCSD(T)-F12}\left(v_i+\frac{1}{2}\right)+\sum_{i\geq j} x_{ij}^{MP2}\left(v_i+\frac{1}{2}\right)\left(v_j+\frac{1}{2}\right) \qquad (3)$$

where the very accurate theoretical procedure RCCSD(T)-F12 is employed to compute the harmonic contributions $\omega_i$ of the vibrational energies, whereas the $x_{ij}$ anharmonic constants are computed from an MP2 full-dimensional anharmonic force field and vibrational second order perturbation theory. $v_i$ and $v_j$ represent the vibrational quanta. VPT2 anharmonic constants $x_{ij}$ are supplied as supplementary material (see Supplementary Material Document No. [46]).

In principle, within the VPT2 description, all the vibrational modes of RDME are infrared active due to the asymmetric character of the minimum energy geometry.

**Table II:** Vibrational fundamentals (in cm$^{-1}$) computed using second order perturbation theory (VPT2).

| mode | ω (RCCSD(T)-F12/AVTZ) | ν[a] | Exp.[b] | assignment |
|---|---|---|---|---|
| $\nu_1$ | 3277.9 | 3160.3 |  | $CH_2$ asym str |
| $\nu_2$ | 3153.7 | 3016.7 |  | $CH_3$ asym str |
| $\nu_3$ | 3125.7 | 3031.5 |  | $CH_2$ sym str |
| $\nu_4$ | 3087.2 | **2974.3** |  | $CH_3$ asym str |
| $\nu_5$ | 3018.7 | **3013.8** |  | $CH_3$ sym str |
| $\nu_6$ | 1511.7 | 1467.7 | 1468.1 | $CH_3$ def |
| $\nu_7$ | 1502.4 | **1482.1** |  | $CH_3$ def+$CH_2$ bend |
| $\nu_8$ | 1498.2 | **1475.1** |  | $CH_3$ def |
| $\nu_9$ | 1462.2 | 1429.7 |  | $CH_3$ def+$CH_2$ bend |
| $\nu_{10}$ | 1295.3 | **1258.4** | 1253.9 | O-$CH_2$ str |
| $\nu_{11}$ | 1262.1 | 1231.6 | 1226.6 | HCO($CH_2$)bend |
| $\nu_{12}$ | 1179.2 | 1153.2 |  | HCO($CH_3$)bend |
| $\nu_{13}$ | 1141.3 | 1119.2 |  | HCO($CH_3$) bend |
| $\nu_{14}$ | 978.9 | 955.0 | 944.4 | O-$CH_3$ str |
| $\nu_{15}$ | 571.9 | **190.9** |  | $CH_2$ wag |
| $\nu_{16}$ | 437.4 | **433.8** |  | COC bend |
| $\nu_{17}$ | 298.5 | **241.0** |  | $CH_2$ twist |
| $\nu_{18}$ | 165.7 | 146.7 |  | $CH_3$ torsion |

a) Estimated using Equation 3; levels displaced by Fermi resonances are, emphasized in boldface; b) Observed Argon matrix infrared wavenumbers [21]

In Table II, the computed transitions are compared with existing experimental data even though the available material is very limited due to the RDME radical character and high reactivity. One of the few existing papers on spectroscopy was authored by Gong and Andrews (2011) [21] who measured the infrared spectrum in solid Argon. The spectrum is characterized by four infrared absorptions at 1468.1, 1253.9, 1226.6, and 944.4 cm$^{-1}$, which were assigned by deuterium substitution and density functional theory. Their values recorded in Ar matrix are in a good agreement with our computed wavenumbers predicted for an isolated molecule.

VPT2 neglects the minimum interconversion effects. For the large amplitude motions responsible for the non-rigidity, a specific theory must be employed as it is described in the next section of this paper. However, if the anharmonic force field is accurate enough, usually the VPT2 algorithms implemented in Gaussian allows to predict the effect of resonances providing valuable initial sets of parameters [47]. Predicted Fermi displacements and the Fermi resonance parameters are supplied as supplementary material (see Supplementary Material Document No. [46]). Within the VPT2 approximation and after considering the Fermi displacements, four vibrational fundamentals are found to lie below 500 cm$^{-1}$: $\nu_{15}$, $\nu_{16}$, $\nu_{17}$, and $\nu_{18}$. They can be assigned to the CH$_2$ wagging, COC bending, CH$_2$ twist, and the methyl torsion, respectively.

Excited VPT2 vibrational energy levels up to 620 cm$^{-1}$ are shown in Table IV. The comparison between the computed and the few experimental data validates the VPT2 theory which fails when high excitation energies of the $\nu_{15}$ CH$_2$ wagging mode are computed. For this mode, anharmonic effects are really overestimated generating too large contributions (i.e. $\omega_{15}$=571.9 cm$^{-1}$ and $\nu_{15}$=190.9 cm$^{-1}$), and in addition, the computed anharmonic overtone is truly inconsistent (2$\omega_{15}$=1143.8 cm$^{-1}$ and 2$\nu_{15}$=-20 cm$^{-}$ [1]). The anharmonic force field components corresponding to $\nu_{15}$ are not well established due to the shape of potential energy surface in the double minimum region.

For the large amplitude motions, Fermi displacements are expected to be very small. This fact validates the two-dimensional model described in the next section. At the MP2/AVTZ level of theory, the $\nu_{15}$ fundamental appears slightly displaced (~ -2 cm$^{-1}$) caused by resonance between the wagging fundamental and the $CH_2$ twist overtone. To evaluate these results, it has to be considered that VPT2 fails when spectroscopic parameters involving wagging mode excitations are computed.

**Torsion-Wagging 2D-model**

On the base of the vibrational energies and the results of the test of Fermi interactions, the $CH_3$ torsional motion and the $CH_2$ wagging can be separated from the remaining vibrational modes. Then, to determine variationally the low energies, the following Hamiltonian can be applied:

$$\hat{H}(\theta,\alpha) = -\sum_{i=1}^{2}\sum_{j=1}^{2}\left(\frac{\partial}{\partial q_i}\right)B_{ij}(\theta,\alpha)\left(\frac{\partial}{\partial q_j}\right) + V^{eff}(\theta,\alpha), q_i, q_j = \theta, \alpha \quad (4)$$

In this equation, $V^{eff}(\theta, \alpha)$ represents the effective potential given by:

$$V^{eff}(\theta,\alpha) = V(\theta,\alpha) + V'(\theta,\alpha) + V^{ZPVE}(\theta,\alpha) \quad (5)$$

where $V(\theta,\alpha)$ is the ab initio two-dimensional potential energy surface (2D-PES) determined from the total electronic energies of a set of selected geometries. For this purpose, a total number of 32 geometries were chosen for eight different values of the $\alpha$ coordinate (0º, 15º, 30º, 45º, 60º, 75º, 130º, 160º) and four values of the H4C3O1C2 dihedral angle (180º, 90º, -90º, 0º). The two additional conformations with $\alpha$=130º,160º were considered to assure a correct fit, although only the six first values of $\alpha$ are needed to describe the shape of surface in the low energy region. In order to take into consideration the neglected vibrational modes, 16 curvilinear internal coordinates were allowed to be relaxed in all the geometries.

The Podolsky pseudopotential V'(θ, α) and the $B_{ij}$(θ, α) kinetic energy parameters were determined from the chosen geometries using the ENEDIM code [32] (see Ref. [33] and [34] for details). $V^{ZPVE}$(θ, α) represents the zero point vibrational energy correction, which was determined at the MP2 level of theory within the harmonic approximation. Previous works show the relevance of this correction for obtaining reliable results [48]. The kinetic parameters, the pseudopotential and the vibrational correction of the potential energy surface are supplied as supplementary material (see Supplementary Material Document No. [46]). The V' pseudopotential is very small and almost negligible.

The energies were fitted to symmetry adapted Fourier series type:

$$V(\theta,\alpha) = A^{00} + \sum_{mn} A_{cc}^{nm}(\cos 3m\theta \cos n\alpha) + \sum_{mn} A_{ss}^{nm}(\sin 3m\theta \sin n\alpha) \qquad (6)$$

The terms V' and $B_{ij}$ of the two dimensional Hamiltonian are expressed with analytical series formally identical to Eq. (6). The parameters of the final effective potential energy surface and the independent coefficients corresponding to the kinetic energy parameters are given in Tables III and IV, respectively.

Table III: RCCSD(T)-F12 coefficients (in cm$^{-1}$) of the effective potential energy surface $V^{eff}$(θ, α) according to the symmetry adapted Fourier series (Eq. 6)

| Terms | | Coefficient | Terms | | Coefficient |
|---|---|---|---|---|---|
| $A^{00}$ | 1 | 4493.072 | $A_{cc}^{61}$ | cos3θ cos6α | -4.026 |
| $A_{cc}^{10}$ | cosα | -3320.696 | $A_{cc}^{02}$ | cos6θ | -0.775 |
| $A_{cc}^{20}$ | cos2α | -2367.847 | $A_{cc}^{12}$ | cos6θ cosα | 0.468 |
| $A_{cc}^{30}$ | cos3α | 522.884 | $A_{cc}^{22}$ | cos6θ cos2α | 0.204 |
| $A_{cc}^{40}$ | cos4α | 2013.802 | $A_{cc}^{32}$ | cos6θ cos3α | 0.773 |
| $A_{cc}^{50}$ | cos5α | -1280.869 | $A_{cc}^{42}$ | cos6θ cos4α | 1.380 |
| $A_{cc}^{60}$ | cos6α | 284.964 | $A_{cc}^{52}$ | cos6θ cos5α | -1.106 |
| $A_{cc}^{01}$ | cos3θ | 219.310 | $A_{cc}^{62}$ | cos6θ cos6α | 0.720 |
| $A_{cc}^{11}$ | cos3θ cosα | 21.732 | $A_{ss}^{11}$ | sin3θ sinα | -111.993 |
| $A_{cc}^{21}$ | cos3θ cos2α | 16.904 | $A_{ss}^{21}$ | sin3θ sin2α | 31.900 |
| $A_{cc}^{31}$ | cos3θ cos3α | 1.445 | $A_{ss}^{31}$ | sin3θ sin3α | -13.787 |
| $A_{cc}^{41}$ | cos3θ cos4α | -9.024 | $A_{ss}^{41}$ | sin3θ sin4α | 20.143 |
| $A_{cc}^{51}$ | cos3θ cos5α | 8.576 | $A_{ss}^{51}$ | sin3θ sin5α | -13.717 |

For the surface, the fit parameters ($R^2$=0.999999 and σ=1.9 cm-1) guarantee the correct shape of the potential. Figure 3 represents the low energy region of the two-dimensional potential energy surface (2D-PES).

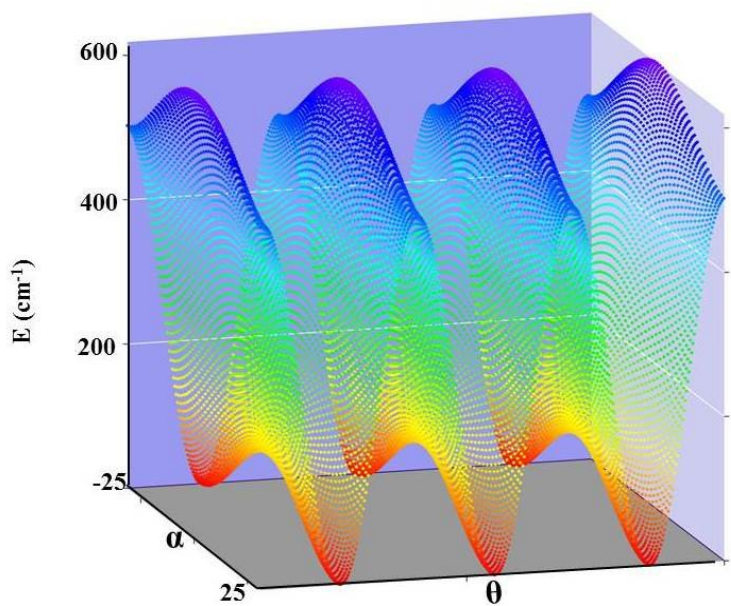

**Figure 3:** The two-dimensional energy surface in the region of the minimum energy configurations.

The 2D-Hamiltonian was solved variationally using ENEDIM [32]. The size of the Hamiltonian matrix was reduced considering the symmetry properties. Details about the procedure and trial functions can be found in references [33-34,40]. The final energy levels $E_{VAR}$, are collected in Table IV. Each level splits into 3 sublevels corresponding to the three methyl torsion minima: one non-degenerate ($A_1$ or $A_2$) and the two-degenerate (E) components. Symmetry is employed for the labelling of the levels as well as, the quantum numbers $n_\theta$, $n_\alpha$.

The barrier heights are also provided in Table IV. The inversion $V^\alpha$ barrier at α=0º was evaluated to be 109 cm$^{-1}$ while the zero point vibrational energy has been obtained to be 280.590 cm$^{-1}$ (the contribution of the wagging mode to the ZPVE is $v_{15}/2 \sim 164$ cm$^{-1}$ > $V^\alpha$). This means that the very low barrier has no effect on the vibrational levels. The $A_i$ and E subcomponents of the levels which splitting is due to the $V_3$ methyl torsion barrier, can be classified as the levels of a molecular species which potential energy surface shows only three minima. One-dimensional curves of Figure 4 can help to understand this problem.

**Table IV:** Low vibrational energy levels of $CH_3$-O-$CH_2$ (cm$^{-1}$)

| $n_\theta, n_\alpha$ | sym. | $E_{VAR}$ | VPT2 | | $n_\theta, n_\alpha$ | sym. | $E_{VAR}$ | VPT2 | |
|---|---|---|---|---|---|---|---|---|---|
| 0 0 | $A_1$ | 0.000[a] | | 0.0 | 0 2 | $A_1$ | 507.547 | $2\nu_{15}$ | ? |
| | E | 0.010 | | | | E | 557.318 | | |
| 1 0 | $A_2$ | 168.376 | $\nu_{18}$ | 146.7 | 2 1 | $A_2$ | 636.522 | $2\nu_{18}\nu_{15}$ | 446.8 |
| | E | 168.013 | | | | E | 636.730 | | |
| | | | $\nu_{17}$ | 241.0 | | | | $2\nu_{17}$ | 482.2 |
| 2 0 | $A_1$ | 311.708 | $2\nu_{18}$ | 290.6 | 4 0 | $A_1$ | 752.051 | $4\nu_{18}$ | 525.4 |
| | E | 316.690 | | | | E | 659.253 | | |
| 0 1 | $A_2$ | 328.557 | $\nu_{15}$ | 190.9 | | | | $3\nu_{18}\nu_{15}$ | 559.3 |
| | E | 328.505 | | | | | | | |
| | | | $\nu_{18}\nu_{17}$ | 378.6 | | | | $2\nu_{17}\nu_{15}$ | 564.9 |
| 3 0 | $A_2$ | 456.258 | $3\nu_{18}$ | 409.4 | | | | $\nu_{18}\nu_{16}$ | 582.0 |
| | E | 423.360 | | | | | | | |
| | | | $\nu_{16}$ | 433.8 | | | | $2\nu_{17}\nu_{18}$ | 601.6 |
| 1 1 | $A_1$ | 486.958 | $\nu_{18}\nu_{15}$ | 323.9 | | | | $3\nu_{18}\nu_{17}$ | 616.8 |
| | E | 496.275 | | | | | | | |
| | | | $\nu_{17}\nu_{15}$ | 385.2 | | | | $\nu_{18}\nu_{15}$ | 619.6 |

| Kinetic energy parameters and effective potential energy barriers (cm$^{-1}$) | | | |
|---|---|---|---|
| $A^{00}(B_{\theta\theta})$ | 7.1432 | $V_3(\alpha=\alpha^{MIN})$ | 502 |
| $A^{00}(B_{\alpha\alpha})$ | 32.0043 | $V_3(\alpha=0º)$ | 510 |
| $A^{00}(B_{\theta\alpha})$ | 1.4246 | $V^\alpha$ | 109 |

a) ZPVE= 280.590 cm$^{-1}$

The methyl torsion barrier was obtained to be ~500 cm$^{-1}$ ($V_3(\alpha=\alpha^{MIN})$=502 cm$^{-1}$; $V_3(\alpha=0º)$= 510 cm$^{-1}$). The barrier is low enough to induce non-negligible splittings. For example, the ground vibrational state splits 0.01 cm$^{-1}$. The lower energies belong to the excited methyl torsional states ($\nu_{18}$) which fundamental level has been found to lie at 168.376 cm$^{-1}$ ($A_2$) and 168.013 cm$^{-1}$ (E). $2\nu_{18}$ was computed to be 311.708 cm$^{-1}$ ($A_1$) and 316.690 cm$^{-1}$ (E). These latter levels lying below the methyl torsion barrier top shows a splitting larger than 5 cm$^{-1}$.

For the methyl torsion, there is a reasonable agreement between the results obtained using the variational procedure and those derived from VPT2. However, as explained in the previous section, very relevant inconsistencies are obtained for the levels assigned to the CH$_2$ wagging mode where the anharmonic contributions derived from VPT2 are entirely unfounded. The fundamental $\nu_{15}$ has been determined to be 328.557 cm$^{-1}$ (A$_i$) and 328.505 cm$^{-1}$ (E) using the variational procedure far away from the result of 190.9 cm$^{-1}$ obtained with VPT2. According to VPT2, the molecule is treated as a semi-rigid species with a single C$_1$ minimum of very low stability given the size of the V$^\alpha$ barrier.

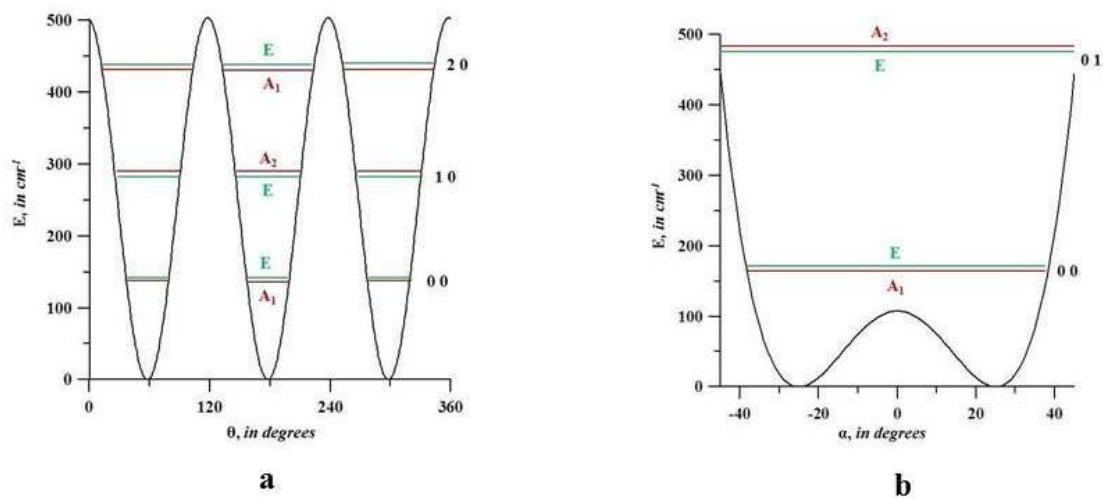

**Figure 4:** Torsional energy levels located in the one-dimensional cuts of the potential energy surface as a function of: a) the θ methyl torsional coordinate, α = α$^{MIN}$; b) the α inversion coordinate, θ = θ$^{MIN}$.

The results of the variational calculations correspond to a molecular structure with a potential energy surface of three equivalent minima of C$_s$ symmetry. The splittings of the levels is due to the methyl group torsion because the CH$_2$ torsion is expected to carry out negligible effects. In addition, the wagging mode has not any effect because the levels lie over the inversion barrier. For further assignments of the experimental spectra, RDME must be treated as a C$_s$ species instead to an asymmetric species. This last structure is not visible in experiments because measurements are not possible bellow the ZPVE.

**CONCLUSIONS**

In this paper, we provide structural and rovibrational parameters for methoxymethyl radical, an undetected relevant astrophysical molecule that has not been previously characterized at the laboratory. The present work aims to help further experimental studies and astrophysical observations.

The ground electronic state of double spin-multiplicity ($X^2A$) appears "clean" from non-adiabatic effects. At the MRCI level of theory, the first excited electronic state, a doublet state, was found to lie at 3.95 eV. The minimum energy geometry is an asymmetric structure which rotational constants have been predicted to be $A_0$=46718.6745 MHz, $B_0$=10748.4182 MHz, and $C_0$=9272.5105 MHz. The dipole moment was determined to be 1.432 D ($\mu_A$=0.6952D, $\mu_B$=1.215D, $\mu_C$=0.3016D).

Three internal motions $CH_2$ wagging, and the $CH_2$ and the methyl torsions can intertransform the 12 minima of the potential energy surface. The interconversion barriers are predicted to be $V^\alpha$ =109 cm$^{-1}$, $V^{CH2}$ ~ 2000 cm$^{-1}$, and $V_3(\alpha=\alpha^{MIN})$=502 cm$^{-1}$. Using VPT2 four anharmonic vibrational fundamentals are found below 500 cm$^{-1}$: $\nu_{15}$, $\nu_{16}$, $\nu_{17}$, and $\nu_{18}$. They can be assigned to the $CH_2$ wagging, COC bending, $CH_2$ torsion, and the methyl torsion, respectively. Since the splittings due to the $CH_2$ torsion are negligible, the motion can be treated as a $CH_2$ twist. By taking into consideration the results of the test of resonances and the ab initio barriers heights, a two-dimensional model depending on the methyl torsion and the $CH_2$ wagging can be suitable. The levels computed variationally splits into 3 sublevels corresponding to the three methyl torsion minima: one non-degenerate ($A_1$ or $A_2$) and the two-degenerate (E) components, band structure that corresponds to a surface with only three minima because the top of the inversion $V^\alpha$ barrier at $\alpha$=0º (109 cm$^{-1}$) stands below the zero point vibrational energy. The $C_1$ minimum energy structure cannot be observed in experiments where RDME appears as a species of $C_s$ symmetry because it stands below the zero point vibrational energy, region inaccessible for experiments.

**ACKNOWLEDGEMENTS**

This research was supported by the **FIS2016-76418-P** project of the "Ministerio de Ciencia, Innovación y Universidades" of Spain and the CSIC i-coop 2018 programme **COOPB20364**. The authors acknowledge the **COST Actions CM1401 "Our Astrochemical History" and CM1405 "MOLIM".** The calculations have been performed in the **CESGA** and **CTI-CSIC** computers centers. MC also acknowledges the financial support by the Consejería de Conocimiento, Investigación y Universidad, Junta de Andalucía and European Regional Development Fund (ERDF), ref. SOMM17/6105/UGR.